# Deep Learning and Bayesian Deep Learning Based Gender Prediction in Multi-Scale Brain Functional Connectivity

Gengyan Zhao[a,*], Gyujoon Hwang[a], Cole J. Cook[a], Fang Liu[b], Mary E. Meyerand[a,b,c], and Rasmus M. Birn[a,d]

[a] Department of Medical Physics, University of Wisconsin – Madison, USA

[b] Department of Radiology, University of Wisconsin – Madison, USA

[c] Department of Biomedical Engineering, University of Wisconsin – Madison, USA

[d] Department of Psychiatry, University of Wisconsin – Madison, USA

[*]Corresponding Author: Gengyan Zhao, Ph.D.,

E-Mail: gzhao23@wisc.edu

# Abstract

Brain gender differences have been known for a long time and are the possible reason for many psychological, psychiatric and behavioral differences between males and females. Predicting genders from brain functional connectivity (FC) can build the relationship between brain activities and gender, and extracting important gender related FC features from the prediction model offers a way to investigate the brain gender difference. Current predictive models applied to gender prediction demonstrate good accuracies, but usually extract individual functional connections instead of connectivity patterns in the whole connectivity matrix as features. In addition, current models often omit the effect of the input brain FC scale on prediction and cannot give any model uncertainty information. Hence, in this study we propose to predict gender from multiple scales of brain FC with deep learning, which can extract full FC patterns as features. We further develop the understanding of the feature extraction mechanism in deep neural network (DNN) and propose a DNN feature ranking method to extract the highly important features based on their contributions to the prediction. Moreover, we apply Bayesian deep learning to the brain FC gender prediction, which as a probabilistic model can not only make accurate predictions but also generate model uncertainty for each prediction. Experiments were done on the high-quality Human Connectome Project S1200 release dataset comprising the resting state functional MRI data of 1003 healthy adults. First, DNN reaches 83.0%, 87.6%, 92.0%, 93.5% and 94.1% accuracies respectively with the FC input derived from 25, 50, 100, 200, 300 independent component analysis (ICA) components. DNN outperforms the conventional machine learning methods on the 25-ICA-component scale FC, but the linear machine learning method catches up as the number of ICA components increases. Second, the proposed feature ranking method was validated with the cross entropy loss on the training dataset. The most important male and female features' prediction accuracy and repeatability in each network structure were studied, and their FC patterns were analyzed and compared with those in previous studies. Third, the behavior of the prediction accuracy and model uncertainty of Bayesian DNN were also studied. Bayesian DNN can maintain the prediction accuracy when Monte Carlo dropout testing is performed, and the model uncertainty increases as the uncertainty within the testing data increases. To test the robustness of the models, all the experiments were done with 50 randomly permuted 2-fold cross validations. Our proposed schemes and the reported findings show the different advantages of the nonlinear DNN and the linear method on different scales of FC inputs, the validity of the proposed feature ranking method and the effectiveness of the model uncertainty generated by Bayesian DNN, and may serve as the very basic framework to verify and study the performance and characteristics of DNN models for further applications to other fields of neuroscience.

# Introduction

The difference in brain structure and function between genders has been a durative topic in the field of neuroscience, and has an important role in determining differences between genders in various psychological and behavioral processes (Gong et al., 2011). For example, it has been known for a long time that males and females are different in memory, language, emotion, perception, navigation and other cognitive categories (Cahill, 2006). At the same time, both structural and functional brain differences between genders have been found in various modalities (Cosgrove et al., 2007; Gong et al., 2011), and brain gender differences widely exist not only in healthy subjects but also in subjects with different kinds of brain disorders, including autism (Alaerts et al., 2016), depression (Orgo et al., 2016) and Alzheimer's disease (Malpetti et al., 2017) etc. Thus, analyzing the brain differences between genders can help improve the understanding of the neurobiological mechanisms behind different gender-related psychological and behavioral processes. Furthermore, for brain disorders that manifest differences between genders, knowledge of brain differences may help to provide a deeper understanding of the psychopathology and aid in the development of related treatments.

## Previous Work

Morphologically, men in general have a slightly larger brain as well as gray and white matter tissue volumes than women (Ruigrok et al., 2014). In addition, the hippocampus, amygdala, neocortex, insula and many other brain regions in charge of different kinds of cognitive processing are sexually dimorphic (Cahill, 2006). (Sowell et al., 2007) showed a difference in cortical thickness between genders across a large age range, and (Lv et al., 2010) reported significant cortical thickening in the frontal, parietal and occipital lobes in women. In diffusion tensor imaging (DTI) studies, females were found having significantly lower fractional anisotropy (FA) in the right deep temporal region and microstructural organization in multiple white matter regions suggested a sexual dimorphism (Hsu et al., 2008), while (Ingalhalikar et al., 2014) illustrated more inter-hemispheric connectivity in females and more intra-hemispheric connectivity in males. (Feis et al., 2013) achieved very high gender prediction accuracy (96%) when using multimodal anatomical and diffusion MR images.

Structural brain differences can help us understand the gender related psychological and behavioral differences to some extent, and functional brain differences can let us move one step further (Gong et al., 2011). Many studies have shown gender differences in FC derived from fMRI in the last decade. Mainly, these studies can be divided into task fMRI (tfMRI) studies and resting state fMRI (rfMRI) studies. (Schmithorst and Holland, 2006) observed a gender-intelligence-age interaction in the FC during the silent verb generation semantic task within a large pediatric group. (Butler et al., 2007) showed the FC difference between men and women in the ventral anterior cingulate cortex and the dorsal anterior cingulate cortex connection during a visuospatial task. In the last decade, rfMRI has drawn enormous attention due to its ability to investigate the spontaneous and intrinsic brain activities. (Bluhm et al., 2008) reported the gender difference in the default mode network. (Biswal et al., 2010) examined the gender difference of resting FC on a large dataset of 1414 subjects collected at 35 international centers, in which men and women showed different connectivity strength in multiple brain connections. Similarly, (Filippi et al., 2013) showed the gender difference between different resting state networks with statistical parametric mapping and independent component analysis (ICA). Recently, (Zhang et al., 2016) used regression and graph theory analyses to show the gender differences in resting FC. For a more comprehensive review, please refer to (Gong et al., 2011; Zhang et al., 2018).

Compared with analyzing the FC group mean differences between genders, extracting important features from an accurate prediction model can disclose the direct relationship between FC and genders. (Casanova et al., 2012) used lasso regression and random forest based methods, and reached 62.3% and 65.4% accuracy respectively on a 148-subject dataset. (S. M. Smith et al., 2013) applied leave-one-out training and testing with multivariate linear discriminant analysis to 104 subjects from Human Connectome Project (HCP) and achieved 87% gender prediction accuracy. Recently, (Zhang et al., 2018) achieved 87% accuracy with partial least squares regression on 820 HCP subjects with 10-fold cross validation. All of these studies also tried to extract the important features from the prediction models to study the FC characteristics of different genders.

In the recent past, tremendous progress has been made in the field of artificial intelligence because of the resurgence of the deep neural network (DNN) methods (Krizhevsky et al., 2012; LeCun et al., 2015) and the rapid advance of parallel computing (Coates et al., 2013; Schmidhuber, 2015). Due to its ability of accurate prediction, DNN has been quickly applied to image processing in neuro-imaging and classification in neuroscience. Convolutional neural network (CNN) based architectures are effective at brain segmentation or skull stripping, while autoencoder, another type of DNN, is broadly used for increasing the prediction accuracy in various neuropsychiatric disorders. (Kim et al., 2016) used autoencoder to classify schizophrenia patients against healthy controls with an accuracy of 85.8% and investigated multiple DNN configurations' effects on the predicting accuracy. Several groups have applied autoencoder based methods to the diagnosis of Alzheimer's disease (Liu et al., 2014; Suk et al., 2015; Hu et al., 2016; Bhatkoti and Paul, 2016) and showed improvement against traditional methods. (Hazlett et al., 2017) used a combination of autoencoder and support vector machine to study the brain development of infants at high risk for autism spectrum disorder. A more comprehensive review can be found in (Vieira et al., 2017).

**Our Approach**

There are two purposes of this work. One is to use deep learning and Bayesian deep learning based methods to perform gender prediction, related feature extraction and model uncertainty generation from the resting state brain FC at multiple scales. The other is to build a framework for testing the reliability, repeatability and robustness of the prediction accuracy, the extracted FC features and the model uncertainty. Gender prediction can also serve as a basic testbed to verify and compare these methods' performance to conventional methods and study the characteristics of these methods for further neuroscience applications.

First, we propose to utilize the highly nonlinear model of DNN to predict gender from brain FC. The trend of the prediction accuracy of DNN at different scales of brain FC was investigated in comparison with that of the linear support vector machine (SVM). Second, instead of treating each connection in the connectivity matrix as a single feature, with DNN the features are extracted as connectivity patterns in the whole connectivity matrix. We propose a method to rank the extracted high-level male and female features based on their contributions to the prediction and studied how much prediction accuracy the several most important features can preserve. Third, Bayesian deep learning was further applied to the FC based gender prediction. The prediction accuracies at multiple dropout rates were compared with the conventional weight averaging technique. The behavior of the model uncertainty generated by Bayesian deep learning for each prediction was also studied. Finally, the repeatability and robustness of all the results were tested through 50 randomly permuted cross validations in all the DNN structures and scales of brain FC studied. All the tests were done on the high-quality large-scale dataset of the Human Connectome Project (HCP) (Van Essen et al., 2013) S1200 release (https://www.humanconnectome.org) at multiple brain connectivity scales derived from different numbers of ICA components. A full comparison was also made between different DNN structures and commonly used machine learning methods.

## Material and Methods

### Dataset and Preprocessing

The rfMRI data of 1003 healthy adults (age: 22~37; 53.2% female) having 4 complete runs in the HCP S1200 release were used for this study (S. Smith et al., 2013). According to the HCP data dictionary, the term "gender" is used instead of "sex" (https://wiki.humanconnectome.org). While "gender" and "sex" are different, participants in the HCP were asked a number of demographic questions, including "gender" (HCP S1200 release reference manual, https://www.humanconnectome.org). We therefore use the term "gender" in this study to conform with the HCP nomenclature. All the data were collected on the customized Siemens Skyra 3T MRI scanner for the HCP with a standard 32-channel Siemens receive head coil. A multi-band gradient-echo EPI sequence was used for the rfMRI with the imaging parameters: TE = 33.1ms, TR = 720ms, Flip Angle = 52°, Multi-band factor = 8. Each subject underwent 4 rfMRI runs in 2 sessions. The duration of each run was 14 minutes 33 seconds, and each run was reconstructed into 1200 3D volumes of 104×90 in-plane matrix size and 2×2 mm$^2$ in-plane pixel size with 72 slices and 2 mm slice thickness. Table 1 is a summary of the demographics, neuropsychological measurements, brain volumes and motion measurements of the male and female groups of subjects involved in this study.

Table 1. Summary of the demographics and measurements of the subjects used.

|  | Male (Mean ± Std) | Female (Mean ± Std) | p-value |
|---|---|---|---|
| **Demographics** |  |  |  |
| Number (%) | 469 (46.76%) | 534 (53.24%) |  |
| Ethnicity (Caucasian/%) | 365 (77.83%) | 393 (73.60%) |  |
| Age (year) | 27.87 ± 3.65 | 29.45 ± 3.61 | <0.001 |
| Education (year)* | 14.87 ± 1.74 | 15.05 ± 1.79 | 0.092 |
| Handedness | 61.04 ± 43.48 | 70.97 ± 43.23 | <0.001 |
| Weight (pound)* | 189.64 ± 34.77 | 155.95 ± 35.50 | <0.001 |
| Blood pressure: systolic (mmHg)* | 128.00 ± 13.25 | 119.40 ± 13.00 | <0.001 |
| Blood pressure: diastolic (mmHg)* | 78.20 ± 10.38 | 74.74 ± 10.46 | <0.001 |
| **Neuropsychological measurement** |  |  |  |
| Fluid intelligence: PMAT24_A_CR* | 17.72 ± 4.59 | 16.41 ± 4.73 | <0.001 |
| Fluid intelligence: PMAT24_A_SI* | 2.42 ± 3.63 | 3.40 ± 3.90 | <0.001 |
| Fluid intelligence: PMAT24_A_RTCR* | 17235 ± 9421 | 14754 ± 8825 | <0.001 |
| **Brain volume** (cm$^3$) (Gray matter + White Matter + CSF) | 1215 ± 96 | 1063 ± 85 | <0.001 |
| **Motion** (mm) (Movement_RelativeRMS_mean) | 0.0862 ± 0.0346 | 0.0876 ± 0.0346 | 0.218 |

* Missing values from some subjects were removed in the calculations.

Handedness: [-100,100], positive numbers indicate more right-handedness; Fluid intelligence: measured with Penn Progressive Matrices (Bilker et al., 2012); PMAT24_A_CR: number of correct responses; PMAT24_A_SI: total skipped items; PMAT24_A_RTCR: median reaction time for correct responses; Brain volume: results from FreeSurfer (http://freesurfer.net/); Motion: temporal mean of the root mean squire of the relative motion, results from HCP minimal preprocessing pipeline (Glasser et al., 2013); for more detailed definitions, please refer to the HCP Data Dictionary for the 1200 Subjects Release (https://wiki.humanconnectome.org).

All the data were preprocessed by the HCP minimal preprocessing pipeline (Glasser et al., 2013) including distortion correction, field map correction, motion correction and spatial normalization. After the gradient distortion correction, head motion of the rfMRI data was corrected by registration to the single-band reference image collected at the beginning of each run. Then the spatial distortion caused by B0 was corrected with field maps, and the single-band reference image was further registered to the T1w structural image. All the preceding transforms were concatenated and applied to the original rfMRI images and the images were resampled into 2mm MNI space. Finally, global intensity normalization was done, and non-brain areas were masked out (S. Smith et al., 2013). The HCP multi-modal surface matching algorithm (MSM-ALL) (Robinson et al., 2014) was used during the preprocessing to improve the inter-subject registration of cerebral cortex with the areal features derived from myelin maps, resting state network and rfMRI visuotopic maps. The artifacts within the rfMRI data were further removed with ICA-FIX (Griffanti et al., 2014; Salimi-Khorshidi et al., 2014) and regression of 24 confound time series derived from motion estimation (including 6 rigid-body parameters, their derivatives and the squares of all the 12).

Spatial group-ICA (Hyvarinen, 1999; Beckmann and Smith, 2004) was applied to the data in the grayordinate space following group principal component analysis (PCA) (Smith et al., 2014) to get ICA based parcellations. Since larger number of components in ICA leads to more and smaller parcels and vice versa, group-ICA was used to decompose the data into several different levels: 25, 50, 100, 200 and 300 components, to analyze brain connectivity at different scales. The associated average blood-oxygenation-level-dependent (BOLD) time series for each subject's ICA components can then be obtained through multiple-spatial-regression of the 4D rfMRI data against the group-ICA spatial maps. Then the brain connectivity matrix for each subject was calculated using the Pearson's correlation coefficient from the concatenated 4 runs' ICA component time series.

The preprocessing steps were performed by the HCP, and the connectivity matrices released in the HCP S1200 Extensively Processed fMRI Data (https://www.humanconnectome.org/study/hcp-young-adult/document/extensively-processed-fmri-data-documentation) were used in this study. Each connectivity matrix was Fisher's r-to-z transformed and normalized to a matrix with zero mean and unit variance to be used as the input of the deep learning classifier. For comparison, the connectivity matrices from the template based parcellation were also calculated. After the ICA-FIX step, the average BOLD time series were extracted from the 330-parcel HCP multi-modal cortical parcellation (Glasser et al., 2016) and the 49-parcel FreeSurfer (http://freesurfer.net/) subcortical parcellation in the grayordinate space (https://balsa.wustl.edu/WK71) (Van Essen et al., 2017). Then, the matrices were Fisher's r-to-z transformed and normalized as above.

## Deep Neural Network

Deep learning is a kind of mathematical model originally inspired by the biological neural system and then widely used for artificial intelligence tasks due to its extraordinary performance (LeCun et al., 2015). As a kind of artificial neural network, deep neural network (DNN) usually contains multiple hidden layers that are fully connected layers. The structure of a typical DNN with three hidden layers is shown in Fig. 1. To simulate the behavior of biological neurons (London and Häusser, 2005), the value of the $k$th neuron in layer $l+1$ of DNN, $A_k^{l+1}$, equals the nonlinearly transformed linear combination of neurons $A^l$ in layer $l$ with a bias $b_k^l$.

The sigmoid function $sgm(x)$ is a commonly used nonlinear transformation in DNN, since it maps a real-valued input to (0,1), which can represent the firing rate of a neuron.

$$A_k^{l+1} = sgm(W_k^{l+1,l} A^l + b_k^{l+1,l})$$
$$A_k^1 = sgm(W_k^{1,0} X + b_k^{1,0})$$
$$sgm(x) = 1/(1 + e^{-x})$$
(1)

$W_k^{l+1,l}$ and $b_k^{l+1,l}$ are the weight matrix and bias, respectively, for layer $l$ pointing to the $k$th neuron in layer $l+1$.

In the last layer $L$, for classification the softmax function is used as the nonlinear transformation to map the real-valued score for each class into normalized estimated class probabilities, and a categorical cross entropy loss function with elastic net regularization of the weights is used to train the DNN.

$$\hat{W}, \hat{b} = \arg \min_{W,b} - \sum_i \sum_k p_i(k) \log q_i(k) + \sum_l \beta^{l+1,l} \left\| W^{l+1,l} \right\|_1 + \sum_l \frac{\gamma^{l+1,l}}{2} \left\| W^{l+1,l} \right\|_2^2$$
$$q(k) = e^{S(k)} / \sum_k e^{S(k)}$$
(2)

$\sum_k p_i(k) \log q_i(k)$ is the cross entropy between the estimated class probability $q(k)$ and the true probability distribution $p(k)$ for the $i$th input in the batch. $\beta$ and $\gamma$ are the weights for the L1 and L2 regularization of the weights in the DNN respectively. $S(k) = W_k^{L,L-1} A^{L-1} + b_k^{L,L-1}$ is the real-valued score for the $k$th class in the last layer. In addition to preventing overfitting, the L1 and L2 normalization is also used based on the assumptions of the human neural system. L2 normalization is used to penalize peaky weight vectors and favor diffuse weight vectors, since we are looking for connectivity patterns for males and females and want the whole connectivity matrix taken into consideration. Meanwhile, L1 normalization is used to drive the weight vector sparse, since only a subset of the connections is assumed to be finally relatively useful for the gender prediction. In addition, a dropout layer is also used following each hidden layer to prevent overfitting.

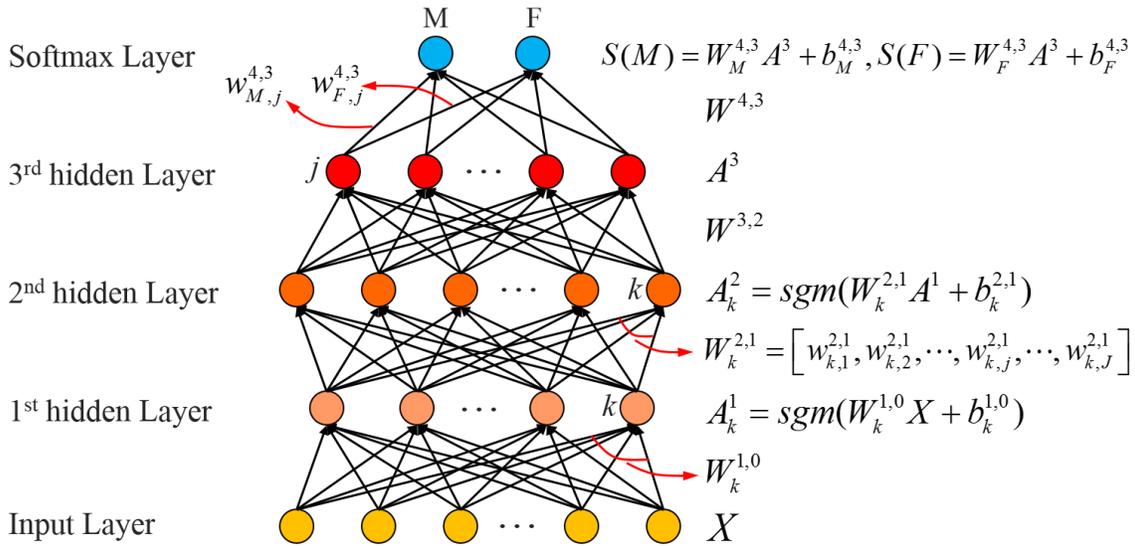

Fig. 1. Illustration of the structure of a typical DNN with 3 hidden layers for classifing 2 classes.

### Deep Neural Network Feature Selection

In the field of neuroscience, knowing which features matter most in a prediction is as important as reaching a high accuracy in the prediction, and at the same time reaching a high accuracy is a necessary condition to make the features selected meaningful. In this study instead of selecting the features based on comparing the group differences of all the input features, we use the framework of selecting features from a high-accuracy prediction model to extract the most important FC patterns from the DNN used for gender classification. Based on the DNN high-level feature extraction approach in the previous study (Kim et al., 2016), we further develop the understanding of the feature extraction mechanism in DNN, and propose a method to rank the features based on their contributions to the final classification. Unlike the feature extraction approaches in other statistical or machine learning models (e.g. logistic regression, SVM etc.), which usually extract individual predictor variables as features that can best distinguish different classes, our method looks for the most important predictor variable patterns in the input for each class as features. Specifically, in this study, our method extracts the most important FC patterns of the whole connectivity matrix for both males and females instead of several individual functional connections that can distinguish the two groups most.

The weight vectors between the input layer and the 1st hidden layer can be considered as convolutional kernels. There are the same number of convolutional kernels as the neurons in the 1st hidden layer, and these kernels are trained to extract the FC patterns which are most helpful on classifying different classes from the input connectivity matrices. $W_k^{1,0}$ is the convolutional kernel pointing to the $k$th neuron in the 1st hidden layer, and only the connectivity matrices $X$ having a similar pattern can make the product $W_k^{1,0}X$ large. $b_k^{1,0}$ is the bias added to the product towards the $k$th 1st-hidden-layer neuron. The biases are also trained to optimize the classification task, and each of them serves as the baseline of the corresponding product between the input and the convolutional kernel. Then, the sigmoid function is used to map the real-value biased convolution result to the activation $A_k^1$, which is in the range (0,1). The more an activation is close to 1, the more the FC pattern in the input connectivity matrix is similar to the trained convolutional kernel's pattern.

The whole set of activations of the 1st hidden layer can be understood as the evaluation on how much each kind of pattern $W_k^{1,0}$ a specific input $X$ has, while between the 1st and 2nd hidden layers the convolutional kernels $W_k^{2,1}$ are used to look for a certain combination of these patterns for the activation $A_k^2$. $w_{k,j}^{2,1}$ is a weight in the convolution kernel $W_k^{2,1}$ pointing to the $k$th neuron in the 2nd hidden layer from the $j$th neuron in the 1st hidden layer. The absolute value of the weight $w_{k,j}^{2,1}$ determines the importance of the corresponding pattern $W_j^{1,0}$, and the sign of $w_{k,j}^{2,1}$ means whether the pattern $W_j^{1,0}$ is preferred to appear in the combination or not. If the absolute value of $w_{k,j}^{2,1}$ is very small, then regardless of whether the corresponding pattern $W_j^{1,0}$ appears or not in the input, the product $w_{k,j}^{2,1}A_j^1$ will not make much difference on the activation $A_k^2$ in the higher layer. Since all the activations are positive values in the range (0,1), a large positive weight $w_{k,j}^{2,1}$ will generate a relatively large positive product $w_{k,j}^{2,1}A_j^1$ and lead the activation $A_k^2$ in the higher layer closer to 1, while a large negative weight value will lead the activation more towards 0. When it comes to the higher hidden layers, the higher level of combinations of the patterns are evaluated for the classification task.

The key part of the DNN structure for extracting and ranking the features for each class in the proposed method is between the last hidden layer and the softmax layer. In the application of gender classification, there are only two classes, male and female. Each activation, $A_j^L$, in the last hidden layer $L$ will be multiplied by the weights $w_{M,j}^{L+1,L}$ and $w_{F,j}^{L+1,L}$ respectively towards the male and female neurons, and the baselines $b_{M,j}^{L+1,L}$ and $b_{F,j}^{L+1,L}$ will also be added to them to get the scores $S(M) = W_M^{L+1,L} A^L + b_M^{L+1,L}$ and $S(F) = W_F^{L+1,L} A^L + b_F^{L+1,L}$. The two scores compete with each other and the input will be classified into the class with the larger score, which means the difference between each $w_{M,j}^{L+1,L}$ and $w_{F,j}^{L+1,L}$ pair determines whether the high-level FC pattern represented by $A_j^L$ is used as a male feature or a female feature, and how important it is for the whole DNN classification model. If the difference $w_{M,j}^{L+1,L} - w_{F,j}^{L+1,L}$ is positive, then the corresponding high-level FC pattern is the feature looked for in a subject's connectivity matrix to identify one as a male and vice versa. Meanwhile, the larger the absolute value of the difference is, the more important feature the corresponding FC pattern is in the DNN model for identifying genders, since compared with other FC patterns, the appearance of this FC pattern in a subject's connectivity matrix contributes more to the total difference between the final scores. Therefore, the absolute value of the difference $\left| w_{M,j}^{L+1,L} - w_{F,j}^{L+1,L} \right|$ can be sorted to rank the importance of the high-level features for both male and female classes. Since a high-level feature $F_k^{l+1}$ can be treated as the combination of lower-level features:

$$F_k^{l+1} = \sum_{j \in J} w_{k,j}^{l+1,l} F_j^l$$
$$F_k^1 = W_k^{1,0}$$

(3)

then any high-level feature can be finally represented by the combination of the FC patterns at the original input level (Kim et al., 2016). $F_k^1$ is the $k$th feature at the original input level, which equals to the $k$th convolutional kernel between the input layer and the 1st hidden layer. $J$ is a set of features and their corresponding weights. It can be the universal set including all the features and weights, or a subset including only the weights having relatively large absolute values by a threshold and the corresponding features.

**Bayesian Deep Learning and Bayesian Deep Neural Network**

Bayesian DNN is implemented by adding dropout layers in the network structure and using dropout in both training and testing (Srivastava et al., 2014; Gal and Ghahramani, 2016). A dropout rate in the range (0,1) is set beforehand for each dropout layer, and the neurons in the preceding layer are randomly set to zero at the dropout rate in every iteration during training or in every forward pass during testing. Dropout training offers Bayesian DNNs extra regularization against overfitting (Srivastava et al., 2014), while dropout testing can be viewed as sampling the posterior distribution of the predicted label probabilities over the Bayesian DNN weights (Gal and Ghahramani, 2016).

In dropout training, given the training dataset $X$ and the label set $Y$, the posterior distribution of the network weights $W$, $p(W|X,Y)$, can be obtained through an approximating distribution, $q(W)$, made with variational inference, and $q(W)$ can be inferred by minimizing the Kullback-Leibler (KL) divergence (Gal and Ghahramani, 2015):

$$KL(q(W) \| p(W|X,Y)) \tag{4}$$

In dropout testing, to predict the label $y^*$ for the data $x^*$, the posterior distribution can be inferred by Monte Carlo (MC) dropout testing. In practice, $T$ times of stochastic forward passes are performed and each time a network weight subset $\hat{W}_t$ is sampled (Gal and Ghahramani, 2015).

$$p(y^*|x^*,X,Y) \approx \int p(y^*|x^*,W)q(W)dW \approx \frac{1}{T}\sum_{t=1}^{T}p(y^*|x^*,\hat{W}_t)$$
$$\hat{W}_t \text{ 燔 } q(W) \tag{5}$$

The mean of the sampled probabilities is used as each label's predicted probability, and the variance of them is used as the model uncertainty on each prediction. The uncertainty generated by MC dropout testing can reflect the confidence of the model on each prediction and is particularly useful for the applicability evaluation of the trained model on the testing dataset. In comparison, the conventional weight averaging testing uses the weights multiplied by the corresponding retaining probabilities (1 - dropout rate) during testing time and can only make predictions (Srivastava et al., 2014; Gal and Ghahramani, 2016).

## **Experiments**

### Gender Prediction from Multi-Scale Functional Connectivity

To understand the relationship between brain functional connectivity and gender, and how the scale of brain connectivity affects the identification of a subject's gender, we used the connectivity matrices derived from different numbers of ICA components: 25, 50, 100, 200, 300 components, as the input for the DNN classifier. In case the structure of the DNN also has a significant influence on the prediction accuracy, the performances of DNNs with different numbers of hidden layers and different numbers of neurons in each layer were also studied. DNNs with 1, 2, 3 hidden layers and 20, 50, 100, 200 neurons in the first hidden layer, and half number of neurons in the following hidden layers were used for training and testing. For example, the DNN with 3 hidden layers and 20 neurons in the $1^{st}$ hidden layer has 10 neurons in the $2^{nd}$ and $3^{rd}$ hidden layers, while the DNN with 1 hidden layer and 20 neurons only has a $1^{st}$ hidden layer with 20 neurons. Since the number of neurons in the following hidden layers controls the number of combined features used for classification, using a smaller number of neurons in the higher hidden layers is based on the assumption that a smaller number of $1^{st}$ hidden layer features' combinations are enough as higher-level features for the classification, and this can dramatically reduce the parameter space of the DNN model. The training and testing were carried out in a two-fold cross validation manner on the 1003 subjects (502 subjects in one fold and 501 in another), and 50 randomized permutations were performed on how to divide the training and testing datasets. A three-way ANOVA using an overdispersed binominal logistic regression was used to test the factors of number of ICA components, number of DNN hidden layers and number of neurons in the DNN layer and their interactions on prediction accuracy. Post hoc testing was also performed following ANOVA to determine the prediction accuracies of which parameter combinations were significantly different from one another.

In comparison with the DNN model, supporter vector machines (SVM) with linear and order-2 polynominal kernels were also studied for gender prediction. To compare with the ICA based functional connectivity, connectivity matrices derived from the template based parcellation, the combination of the 330-parcel HCP multi-modal cortical parcellation (Glasser et al., 2016) and the 49-parcel FreeSurfer (http://freesurfer.net/) subcortical parcellation in the grayordinate space (https://balsa.wustl.edu/WK71) (Van Essen et al., 2017), were also used to predict gender. 50 randomized permutations of 2-fold cross validations were also done respectively for these comparative methods, and related statistical tests were also performed.

All the DNNs with different structures were trained with the Stochastic Gradient Descent (SGD) algorithm. The learning rate varies from 0.05 to 0.8 and number of iterations varies from 40 to 300, depending on the DNN structure and the size of the input vector (the number of input FC connections in different numbers of ICA components) to make all the training procedures converge around the loss of 0.1. $\beta$ and $\gamma$ were set as $10^{-6}$ and $10^{-4}$ respectively while a dropout rate of 0.2 was set for the last hidden layer in all the DNN models. All DNN models were implemented in Python with the libraries Keras and Tensorflow, and the SVM models were implemented with the package scikit-learn. All the training and testing were performed on a workstation hosting 2 Intel Xeon(R) E5-2620 v4 CPUs (8 cores, 16 threads @2.10GHz) with 64 GB DDR4 RAM and two GPUs: an Nvidia GTX980Ti GPU with 6 GB memory and an Nvidia TITAN Xp GPU with 12 GB memory. A 64-bit Linux operation system ran on the workstation.

<u>DNN Feature Selection and Robustness Evaluation</u>

After the gender prediction and related statistical analysis, we also extracted and ranked the FC features for gender classification with the proposed method, studied their robustness and analyzed the feature differences between males and females. First, to verify the proposed feature-ranking method, we used several individual high-level male and female feature pairs ranked at different importance levels to make predictions on the training dataset itself, and compared the cross entropy loss achieved by each of these feature pairs. Second, we studied the relationship between the prediction accuracy and the number of learned important feature pairs involved in the prediction by investigating how much prediction accuracy the DNN model can preserve by using only a few most important (only the several highly important) high-level feature pairs during prediction. Third, the repeatability of the features learned from each DNN structure was also studied. All these experiments were done on all the DNN structures and all the numbers of ICA components in all the 50 randomly permuted cross validations. Finally, based on the results of these experiments, the extracted FC features from the selected DNN models were plotted to visualize the differences between males and females.

Bayesian Deep Learning and Monte Carlo Dropout Testing

To study the performance of Bayesian deep learning, MC dropout was performed on the previously trained 3-hidden-layer DNNs with the 3$^{rd}$ hidden layer dropped out in dropout testing. The prediction accuracies of several different dropout rates were studied, and the prediction accuracy of weight averaging testing was also compared. The behavior of the uncertainty generated by Bayesian deep learning on the application of image segmentation was already studied in previous research (Zhao et al., 2018). To verify the validity of the model uncertainty generated by Bayesian DNN in the classification application, in this study we also investigated the model uncertainty's behavior by making tests on the data generated with varying levels of male/female uncertainty. In the testing dataset of each cross validation fold, 200 female subjects and 200 male subjects were randomly selected and referred to as the female subset and the male subset. Then, 200 0.75-female-0.25-male subjects were made up by linearly combining the connectivity matrices in the female and male subsets and referred to as the 0.75-female-0.25-male subset. Each subject's connectivity matrix in the 0.75-female-0.25-male subset is the sum of 0.75 times a random connectivity matrix from the female subset and 0.25 times a random connectivity matrix from the male subset, and each connectivity matrix in the female and male subsets was used only once. In a similar manner, the 0.5-female-0.5-male subset and the 0.25-female-0.75-male subset were also created. In each cross validation, the prediction accuracy and corresponding model uncertainty on all the 5 subsets were generated by the Bayesian DNNs with the 3rd hidden layer followed by a dropout layer. Based on the previous experiment, to keep both the prediction accuracy and enough variation a dropout rate of 0.5 was selected. Since the Bayesian deep learning models are trained with only the real female and male subjects, and the made-up subjects have more uncertainty in the data themselves, on the made-up subsets the prediction accuracies of the Bayesian DNNs are supposed to be lower and the corresponding model uncertainties are supposed to be higher.

# Results

## Gender Prediction from Multi-Scale Functional Connectivity

The performances of DNN with various model structures on gender prediction at different scales of brain FC were evaluated. In addition, the SVMs with linear kernel and order-2 polynomial kernel were also compared. In Fig. 2 the mean and standard deviation of the accuracies across the 50 randomized cross validation permutations are plotted. The accuracy in each permutation is the prediction accuracy on the 1003 HCP subjects evaluated from the 2-fold cross validation. As is seen in Fig. 2, as the number of ICA components increases, the scale of the brain FC under investigation becomes finer, and the gender prediction accuracy increases and gradually plateaus. For the predictions from the same scale of FC but different DNN structures, the prediction accuracies are relatively close. The accuracy of the SVM prediction also goes up and gradually plateaus as the FC scale becomes more and more detailed. The order-2 polynomial SVM is slightly better than the linear SVM for the case of 25-component ICA, but for larger numbers of ICA components the linear SVM is better than the order-2 polynomial SVM. In the comparison between DNN and SVM, DNNs' predictions are better than the SVMs predictions for the 25-component ICA, and as the number of ICA components increase the prediction accuracy of the linear SVM catches up: DNNs are only slightly more accurate than linear SVM for 50-component ICA, and linear SVM is better than the DNNs for the FC from larger number ICA components. The prediction accuracies for DNNs and SVMs on the brain FC derived from the template based parcellation were also shown in Fig. 2 for reference. Although there are 330 cortical parcels and 49 subcortical parcels in total, the prediction accuracies are much smaller than the predictions for the 100-component ICA. All the exact mean and standard deviation values in Fig. 2 are shown in the Supplementary Table S1 and S2.

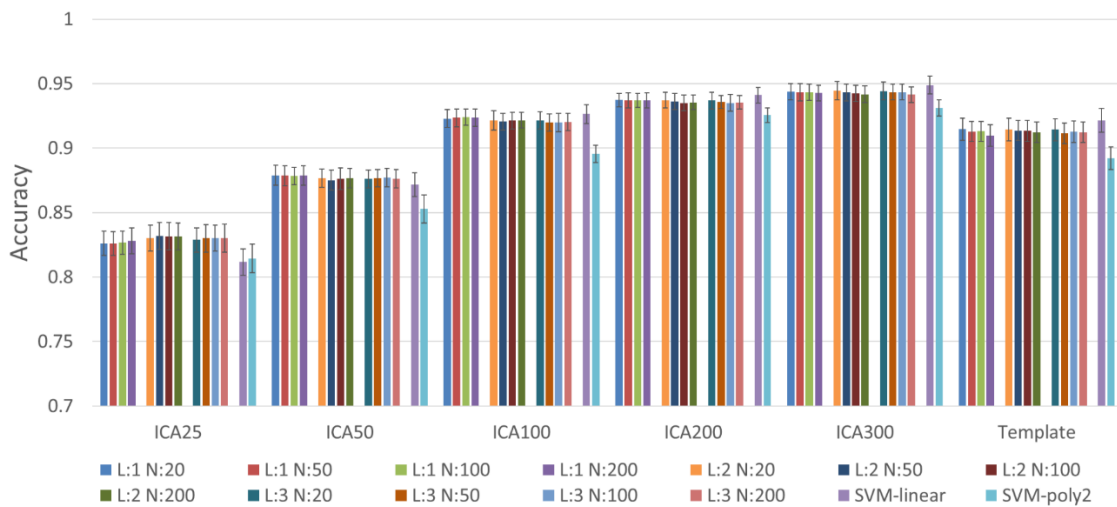

Fig. 2. The mean and standard deviation of the prediction accuracies across the 50 cross validation permutations for each kind of predictive model and each kind of input. L – number of hidden layers in the DNN model; N – number of neurons in the first hidden layer of the DNN, the number of neurons in each of the rest hidden layers is half of this number.

The corresponding statistical tests were also carried out to show the significance of the differences between the reported accuracies. First, a three-way ANOVA test using an overdispersed binomial model was performed on the accuracies generated by DNNs on different scales of FC derived from different numbers of ICA components. Due to the inclusion of interactions in the model, type II ANOVA was used on the 3 factors: the ICA component number, the DNN hidden layer number and the DNN neuron number. The ANOVA test was performed in a hierarchical manner with the full model including all the interactions at the beginning. The factor of ICA component number is significant with a p-value $\ll 0.001$, and interaction between the ICA component number and layer number is significant with a p-value $\ll 0.001$. Other factors and interactions are not significant at the 0.05 significance level. Multiple comparison was done for the significant interaction in the post hoc analysis with the Tukey adjustment. For all the 105 pairwise comparisons, the differences between different ICA component numbers are significant with Tukey corrected p-values $< 0.01$, regardless of the other factors. The differences within the group having the same ICA component number are mainly insignificant at the 0.05 significance level, except that in the 25-component ICA group the difference between the 2-layer and 1-layer DNNs is significant (Tukey corrected p-value < 0.01), and in the 100-component ICA group the difference between 2-layer and 1-layer DNNs (Tukey corrected p-value = 0.0366) and the difference between 3-layer and 1-layer DNNs (Tukey corrected p-value < 0.01) are also significant. Second, another two-way ANOVA test using an overdispersed binomial model was carried out to compare the 1-hidden-layer DNN with 20 neurons against the linear SVM across all numbers of ICA components. Type II ANOVA was used on the 2 factors of the ICA component number and the prediction method (1-hidden-layer 2-neuron DNN vs. linear SVM). Both factors and the interaction are significant (p-values << 0.001). To investigate the significance of the difference between DNN and linear SVM at each FC scale, uncorrected post hoc testing p-values were reported in Table 2. With the conservative Bonferroni correction, the difference for the 100-component ICA is insignificant at the 0.05 significance level, and difference for the 200-component ICA is also insignificant at the 0.01 significance level. The mean difference and the p-value in the table also show that, for small number of ICA components (large scale FC) the DNN is more accurate than linear SVM, and as the number of ICA components increases, the FC becomes more detailed, and the prediction accuracy of linear SVM catches up and finally surpasses the prediction accuracy of DNN.

Table 2. Multiple comparisons between DNN and linear SVM at each FC scale.

| ICA Components | Prediction Accuracy Mean Difference (DNN-SVM) | P-value (uncorrected) |
|---|---|---|
| 25 | 1.456E-02 | 5.214E-13 |
| 50 | 7.298E-03 | 2.447E-05 |
| 100 | -3.490E-03 | 1.157E-02 |
| 200 | -3.829E-03 | 2.215E-03 |
| 300 | -5.184E-03 | 1.127E-05 |

## DNN Feature Selection and Robustness Evaluation

The proposed feature ranking method is verified by the comparison of the cross entropy loss obtained from each feature pair in predicting the training data. The male and female features in the last hidden layer of each DNN were extracted and ranked by their contributions to the difference between the final male and female scores. Then, predictions were made with the 1st, 2nd, 3rd, 4th and 5th most highly ranked male-female feature pair in each DNN respectively on the training data themselves. The cross entropy loss achieved by each of these high-level feature pairs for each DNN structure and each number of ICA components in all the 50 randomly permuted cross validations the are shown in Fig. 3. In all the situations, as the rank of the feature pair goes higher, the lower cross entropy loss it can get when predicting the training data. This means the more highly ranked feature is better at making the predicted label distribution similar to the ground truth distribution and, therefore, the importance and contribution of the features ranked by the proposed method in the DNN model are verified.

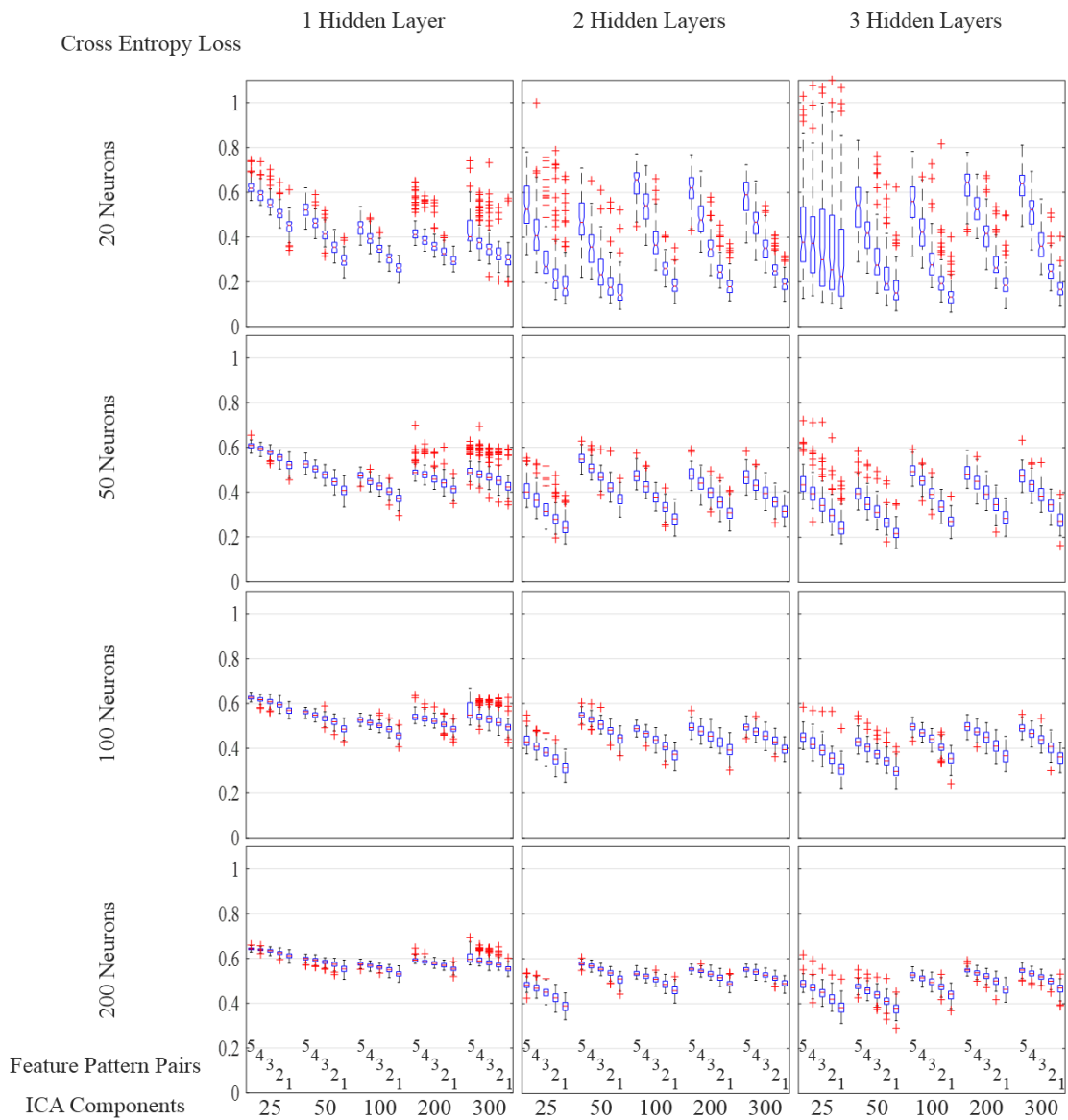

The relationship between the prediction accuracy and number of learned features used in the prediction was also studied, and the results are shown in Fig. 4. For each kind of DNN structure and number of ICA components, only a subset of the most important male and female features in the last hidden layer learned from the training dataset were kept in the DNN model for the prediction on the testing dataset. The most important 1, 2, 5 and 10 male and female feature pairs, which have the largest weight differences towards the final male and female neurons were used for each prediction in all the 50 randomly permuted cross validations, and the resulting accuracies were compared with those from the predictions with all the feature pairs. As is shown in Fig. 4, in general for all the DNN structures and numbers of ICA components, as the number of important feature pairs used in the prediction increases accumulatively, the more accuracy the DNN can recover compared to the prediction made by all the features. This is because all the features in a certain DNN are trained to work together to reach the highest prediction accuracy (lowest loss), and the more complete the DNN model is, the more accuracy can be preserved. It is shown that in Fig. 4 several predictions with fewer features have slightly higher accuracy than the predictions made by more features. This could be caused by the slight inconsistency of the internal features between the training and testing datasets. In the Supplementary Fig. S1 and S2, predictions were also made with the same subsets of learned features on the training dataset. As the number of features used increases, the prediction accuracy increases monotonically and the prediction loss decreases monotonically. Fig. 4 also shows that in the task of predicting gender from brain FC, the most important feature pair can usually recover the majority of the prediction accuracy achieved by all the features. (For example, for the 1-hidden-layer 20-neuron DNN at the scale of 25 ICA components, a median accuracy of 0.795 is achieved by using only the single most important feature pair vs. a median accuracy of 0.827 when using all the features.)

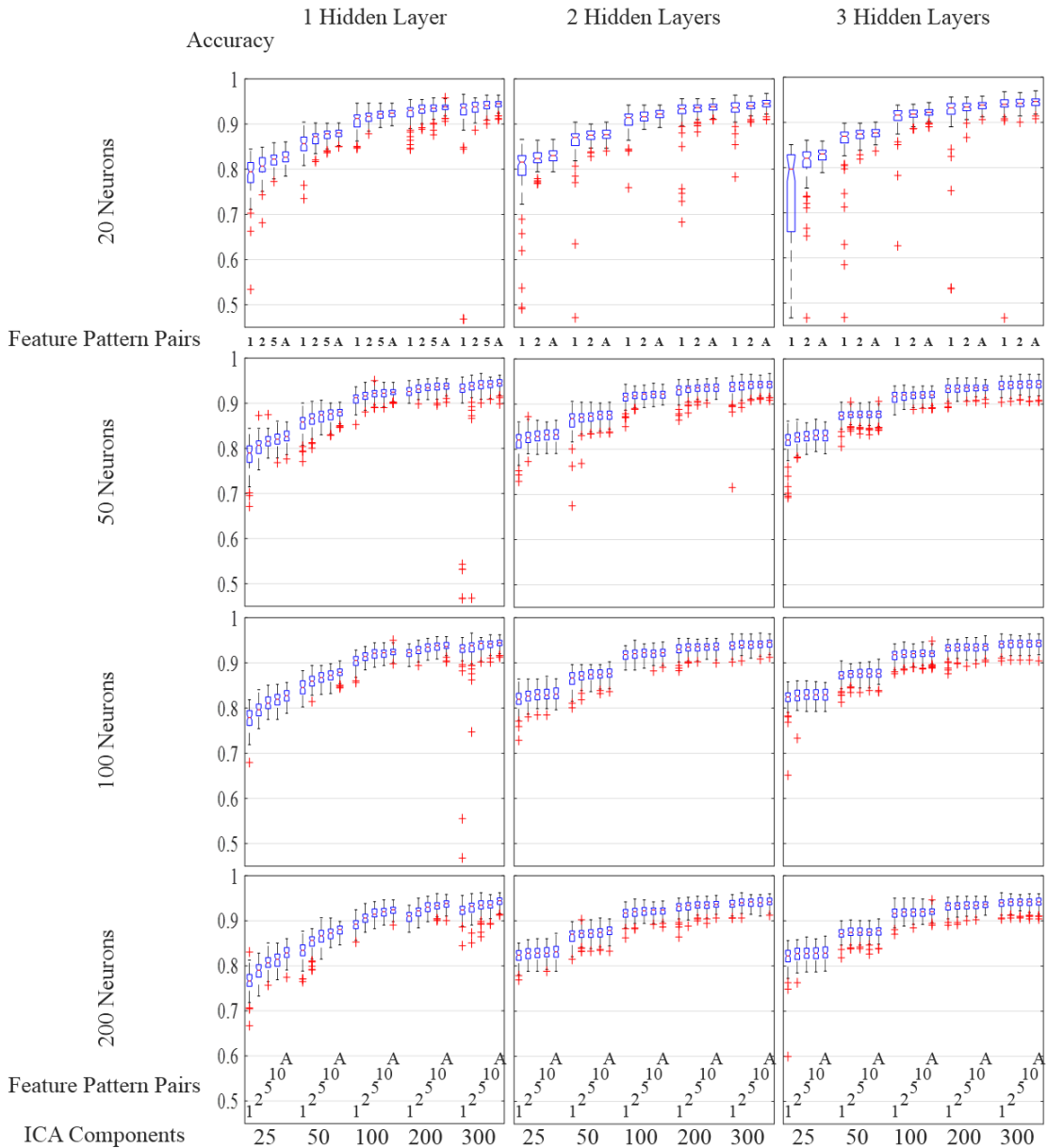

Fig. 4. Prediction accuracy recovered by the several most important high-level male and female feature pairs in the predicitons on the testing dataset. '1', '2', '5' and '10' mean that the predictions were made by the most important 1, 2, 5, 10 male and female feature pairs in the last hidden layer respectively. 'A' means the predictions were made by all the features.

To study the robustness and repeatability of the most important high-level male and female features extracted by different neural network structures for each FC scale, the correlations between the features across all the 50 randomly permuted cross validations were calculated and are shown in Fig. 5. As the neural network gets deeper, the correlation becomes higher, which means that the repeatability of the most important high-level feature is higher. This is true for all the FC scales and numbers of network neurons, although for the 25-ICA-component FC, the differences in correlations from a 2-hidden-layer networks compared to a 3-hidden layer networks are minimal. In addition, as the number of ICA components increases, the correlation generally decreases.

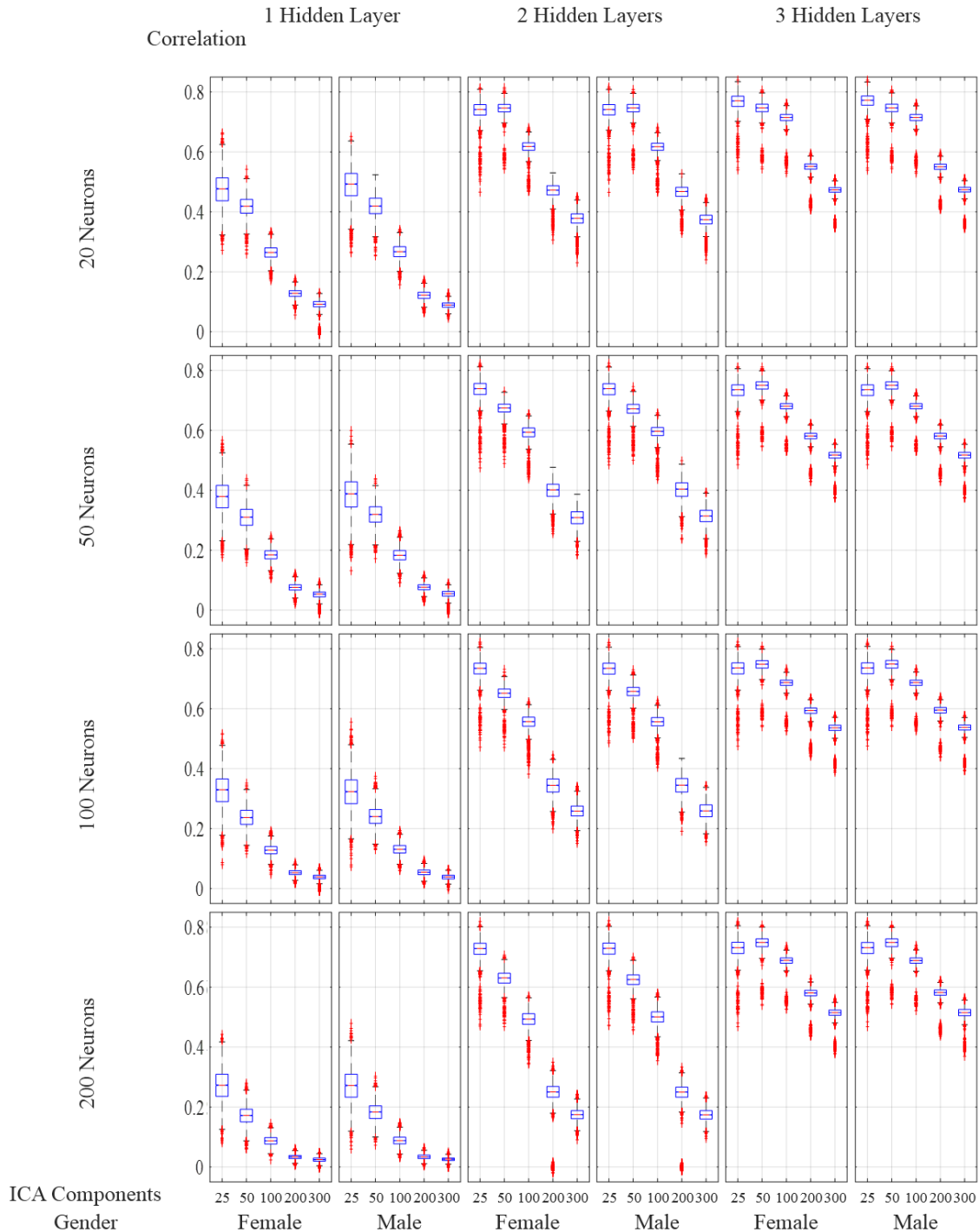

Fig. 5. Correlations of the most important high-level features across all the 50 randomly permuted cross validations for each neural network structure, number of ICA component and gender. For each number of neuron, hidden layer and ICA component, male and female features in the highest level hidden layer were extracted and ranked with the proposed method. The correlations of the highliest ranked features were calculated across the 50 randomly permuted 2-fold cross valadtions (each boxplot shows the correlations between 100 extracted features).

Considering both the accuracy that can be preserved by the most important feature pairs and the repeatability of these feature pairs in the randomly permuted cross validations, the most important high-level brain FC feature pairs extracted by the 200-neuron 2-hidden-layer DNN from the 25-component ICA connectivity matrices and those by the 200-neuron 3-hidden-layer DNN from the 50-component ICA connectivity matrices are shown in Fig. 6. First, from Fig. 6 it can be seen that the group means of the male and female connectivity matrices are similar, and the connectivity patterns of the group means are basically consistent for different number of ICA components. Second, the extracted high-level features by DNN of different genders basically have reversed patterns. This is because DNN is trained to discriminate these 2 genders by looking for the relative pattern difference in their connectivity inputs. This is true regardless of the number of ICA components. Third, the difference of the extracted most important high-level features between genders captures some of the characteristic patterns in input group mean difference, but does not exactly match the input group mean difference. The feature extracted by DNN is trained to maximize the gender prediction accuracy for every subject, and the final high prediction accuracy achieved by DNN is by comprehensively combining all the learned features in training to make predictions.

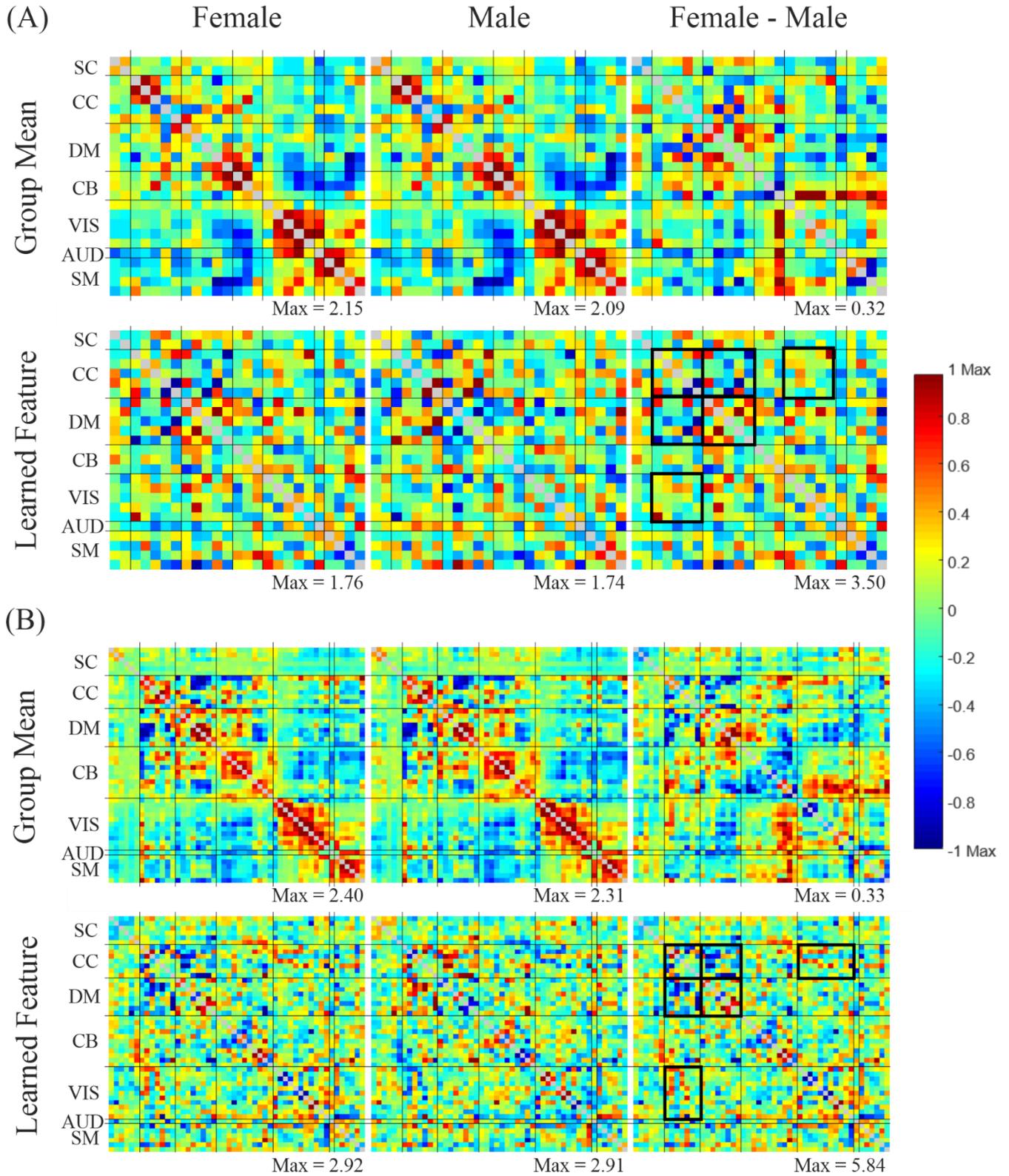

Fig. 6. The most important high-level brain FC feature pairs extracted by DNN. (A) DNN: 200-neuron 2-hidden-layer network; input: 25-component ICA connectivity matrices. (B) DNN: 200-neuron 3-hidden-layer network; input: 50-component ICA connectivity matrices. Each group mean of the input is also shown for comparison. SC, subcortical; CC, cognitive control; DM, default-mode; CB, cerebellar; VIS, visual; AUD, auditory; SM, somatomotor.

In Fig. 6 the most important male and female features extracted by DNN are the most highly ranked high-level FC patterns DNN looks for to make accuracy predictions. For both 25 and 50 ICA components, females have stronger DM-DM, CC-VIS connections, relatively weaker CC-DM connections, and both stronger and weaker connections in CC-CC (in the black boxes in Fig. 6). These findings are consistent with the previous studies (S. M. Smith et al., 2013; Zhang et al., 2018). In Fig. 6 the most important female-male feature differences learned by DNN also show some different patterns for these 2 different FC input scales, for example: the 50-ICA-component feature difference shows stronger and weaker female connections in CB-CB, VIS-VIS, CB-CC, VIS-CC and SM-VIS, whereas the 25-ICA-component feature difference shows stronger female connections in VIS-CC, VIS-DM and a weaker female connection in SM-CC.

## Monte Carlo Dropout Testing and Bayesian Deep Learning

The prediction accuracies of Bayesian deep learning with MC dropout testing at the dropout rates: 0.2, 0.5, 0.9 and the extreme dropout rate R2, which only retains 2 neurons, were studied and compared with the prediction accuracy of weight averaging testing. The results in Fig. 7 show that in general the accuracies achieved by MC dropout testing at low dropout rates (0.2, 0.5) are comparable with the accuracy of weight averaging testing on gender classification. As the dropout rate goes up to extreme large values (0.9, R2), the accuracy of MC dropout testing tends to decrease. The decreasing is especially obvious for the Bayesian DNN with large number of neurons at the dropout rate R2.

Further, the behavior of the uncertainty generated by Bayesian deep learning was also studied by performing MC dropout on the real and made up subsets in dropout testing. Fig. 8 shows that for the 3-hidden-layer Bayesian DNNs trained with the real female and male data, as the uncertainty within the testing data increases, the prediction accuracy decreases and the corresponding uncertainty increases. This is true for all the numbers of neurons and all the numbers of ICA components.

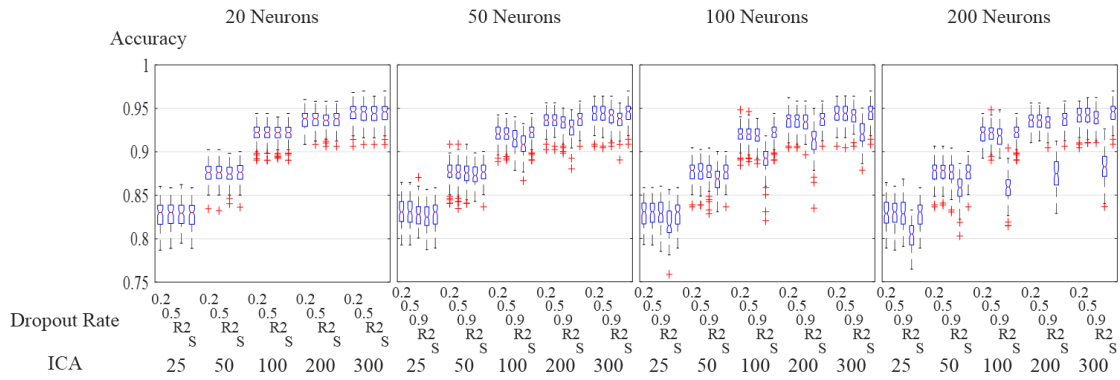

Fig. 7. Prediction accuracy VS dropout rate of MC dropout testing in 3-hidden-layer Bayesian DNNs with the 3rd hidden layer dropped out in all the 50 randomly permuted cross validations for each number of neurons and each number of ICA components. R2 means the dropout rate was set to only retaining 2 neurons; S stands for the result by weight averaging technique during testing. The networks used are previously trained 3-hidden-layer DNNs with the dropout rate of 0.2 on the 3rd hidden layers in dropout training.

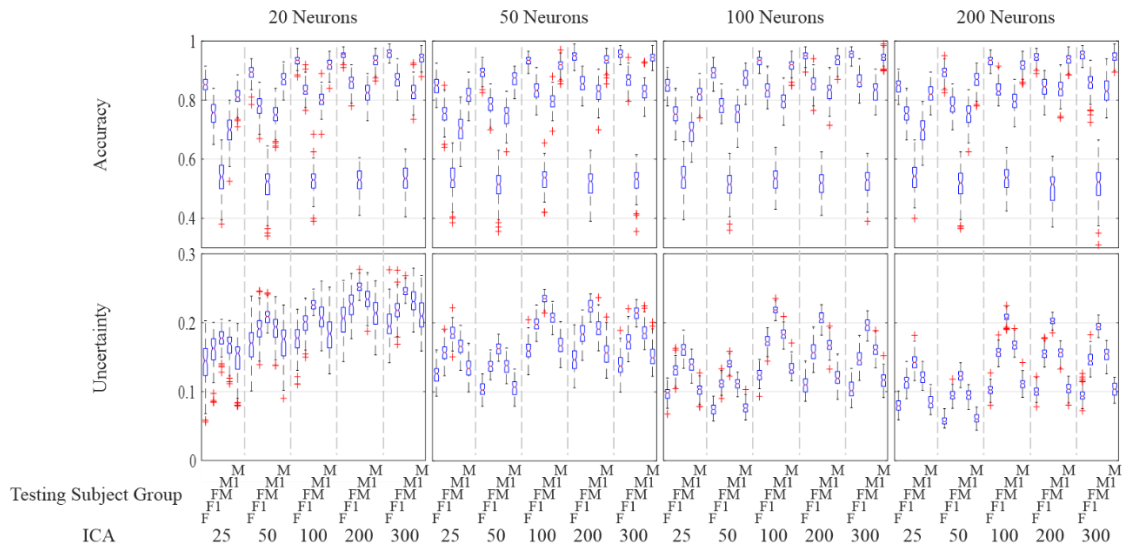

Fig. 8. Prediction accuracy and model uncertainty of Bayesian DNN with MC dropout testing on different testing subsets in all the 50 randomly permuted cross validations for each number of neurons and each number of ICA components. F – real female subsets; F1 – 0.75-female-0.25-male subsets; FM – 0.5-female-0.5-male subsets; M1 – 0.25-female-0.75-male subsets; M – real male subsets. For the purpose of calculation, the labels of the FM subsets were set as female. Previously trained 3-hidden-layer DNNs were used for this experiment, and in MC dropout testing 0.5 dropout rate was set on the 3rd hidden layers.

## Discussion

In this present study, deep learning based methods are applied to the task of gender prediction from resting state brain functional connectivity. The classification performance of DNN was evaluated with different network depths and different number of neurons at different scales of brain FC. The prediction accuracy of DNN was also compared with that of SVMs at different brain FC scales. In addition, based on a new understanding of the DNN model a new feature ranking method was proposed to extract the highly important FC features for the prediction and build the relationship between brain FC and gender. The robustness and repeatability of the features of different network structures and scales of brain FC were also studied. Finally, Bayesian deep learning was applied to the brain FC gender prediction. The prediction accuracy with different dropout rates in MC dropout testing and the behavior of the uncertainty generated by Bayesian DNN were also studied at different brain FC scales. In summary, the experiments on the 1003-subject HCP dataset suggest that the results are highly related to the scale of brain FC under investigation.

Since gender is one of the very basic physiological and psychological properties of human beings (Gong et al., 2011), gender prediction from brain FC on the large-scale high-quality HCP data can serve as a very basic case to verify and study the performance and characteristics of DNN for further neuroscience applications. From the prediction results in Fig. 2 and the following statistical analysis, it can be seen that the results from DNNs with different depths and different number of neurons are very similar, which is not consistent with the application of DNN on classifying brain disorders in previous studies (Kim et al., 2016). This suggests that the performance of DNN in terms of different network structures may be application oriented. In the comparison between DNN and linear SVM, the results show that DNN is much more accurate than linear SVM when the number of ICA components is small. As the number of ICA components increases the accuracy of linear SVM catches up, and when the number of ICA components is relatively large, linear SVM is better than DNN. This illustrates that the performance of the nonlinear DNN versus linear model highly depends on the scale of the FC input. The possible reason for this is the change of the amount of signal and noise in the FC input. When the number of ICA components is small, the signal and noise in the FC is highly reduced, and in this situation the nonlinear DNN model is better at taking the limited signal for accurate prediction. When the number of ICA components is large, the signal is highly redundant but the noise is also greater, and in this case the nonlinear DNN model has no advantage over the linear model. This trend about the linear and nonlinear methods can also be verified through the comparison between the linear SVM and the nonlinear (order 2 polynomial) SVM. Therefore, this study shows that no model is the best for every situation, and that for predictions using brain FC as input the scale of brain FC should be taken into consideration. In the comparison between the ICA based FC and the template based FC, the ICA based FC can reach a better result with a smaller number of components, which means in terms of the increment of signal noise ratio ICA is more efficient than the template based parcellation.

In the proposed feature ranking method, the importance of the extracted high-level features were ranked by their contributions to the difference between the male score and the female score in the softmax layer. Since the cross entropy loss in a prediction reflects how the predicted probability distribution over the classes is similar to the ground truth probability distribution, experiments were performed to compare the cross entropy loss achieved by each high-level feature pair on the training data. Fig. 3 shows that a more highly ranked high-level feature pair can reach a lower cross entropy loss for various DNN structures and inputs, which proves that the importance and contribution of the more highly ranked feature pair is higher. This result verifies the effectiveness and robustness of the proposed ranking method.

Fig. 4 shows as the number of the important high-level features involved in the DNN model increases accumulatively according to their ranks, the prediction accuracy that can be recovered towards the full DNN model also increases. For the application of FC gender prediction, the most important feature pair in various network structures can recover the majority of the accuracy achieved by the full DNN model for all scales of FC input. This is especially true for the networks with more layers and more neurons, which means the redundancy levels of the high-level features in these networks are relatively high. Thus, the ranking method offers us a way to focus on the most important feature pairs in the application of FC gender prediction instead of checking all the hundreds of features learned by DNN. Also, Fig. 4 shows with more lower-level features the most important high-level feature pair in a deep network can recover the accuracy of the whole DNN model better. For the 2-hidden-layer and 3-hidden-layer DNNs, there are large improvements of prediction accuracies recovered with the most important feature pair from 20-neuron DNNs to 50-neuron DNNs.

However, the improvement is not as clear for further increasing the number of low-level features (from 50 to 100 and 200 neurons). This is likely because for a specific application, there is a certain number of useful lower-level features making the main contribution to the final prediction accuracy, and each high-level feature is just a combination of all the low-level features. When the number of lower-level neurons is not enough, increasing it will let a high-level feature combine more useful low-level features to make more accurate predictions. After the certain number is reached, adding more low-level features only increases the redundancy of the network and does not result in much increase in the prediction accuracy recovered by a high-level feature.

The features extracted by DNN during gender classification can only be generalized as the features for brain FC gender classification if they are highly robust and repeatable to different subject groups, otherwise they are more reflective of the features for a specific sample group rather than the features for the whole population. Thus, we investigated the repeatability of the most important feature pairs in all the randomly permuted cross validations for each kind of DNN structure and input FC scale. Fig. 5 shows that as the network gets deeper, the correlation of the most important high-level feature pair gets higher, which means the repeatability of it gets higher. This is because in DNN features of the input are learned hierarchically by hidden layers of different depths, and each high-level feature learned by a deeper hidden layer is constituted of many different low-level features (Lee et al., 2008, 2009; Kim et al., 2016). Since the number of classes to predict is relatively small, in real world applications the number of distinct low-level features needed are much more than that of distinct high level features to make accurate predictions. Thus, the most important high-level features across different training group are more likely to be similar. Also, as the number of ICA components increases, the repeatability of the most important high-level feature generally decreases. This could be because as the dimension of the input increases, more information is contained in the data, and more detailed differences among different training datasets can be described by the high-dimensional input. This also means for the purpose of extracting gender related features, it is better to use relatively smaller number of ICA components.

Although for the application of FC gender prediction DNNs of different depths can achieve similar accuracies in Fig. 2, Fig. 4 and 5 show that deeper network structures still have advantages on the accuracy recovery and repeatability of the most important feature pair. Based on the those results in Fig. 4 and 5, the most important male and female high-level features from two selected DNNs and brain FC scales are shown in Fig 6. First, it can be seen that the characteristic patterns presented by the most important female-male feature differences in both FC scales have similar parts. Both of them captured the relatively high DM-DM and CC-VIS connections, relatively low CC-DM connections and both high and low connections in CC-CC as the female FC features, which means these are very important features for the FC gender classification. These patterns are consistent with the results from previous studies (S. M. Smith et al., 2013; Zhang et al., 2018). Also, the most important female-male feature differences are not all the same across these 2 FC scales, and there are several possible reasons: the input FC group mean patterns are not all the same across these 2 FC scales; different parts of FC can have different importance in the predictions at different FC scales. To achieve the highest prediction accuracy, all the learned features should be combined together to make the decision, since all the features are trained to work together to achieve the best performance. In addition, in Fig. 6 the female-male feature differences extracted by DNN capture some of the patterns in the female-male group mean differences, but not all the large group mean differences. This is because the features extracted by the DNN are the highly important ones for the model to make accurate classification, but not all the large group mean differences are important for the classification, since no standard deviation information is reflected in the group mean differences.

Bayesian DNN can not only make predictions but also report the model uncertainty on each prediction. The prediction accuracies of Bayesian DNN at different dropout rates in Fig. 7 show that in the application of FC gender prediction Bayesian DNN with MC dropout testing can reach the same level of prediction accuracy as the conventional DNN with weight averaging testing as long as the dropout rate is not too low. Thus, 0.5 dropout rate was selected based on the tradeoff between accuracy and variation to test the behavior of the uncertainty generated. As is seen in Fig. 8, the model uncertainty increases as the uncertainty within the testing data increases, which matches the expectation. Since the uncertainty generated by Bayesian DNN can help with picking up the testing subjects whose predictions the model is not very confident about based on the knowledge learned from the training data. In real neuroimaging and neuroscience predication practices, such as classifying the Alzheimer's disease, locating the source of epilepsy, or predicting the effect of a treatment, we can be aware of the uncertainty level of each prediction, and start other procedures to consolidate the highly uncertain prediction results.

There are also some limitations in our study. Based on the results from the previous studies (S. M. Smith et al., 2013; Zhang et al., 2018), we did not regress out the potential confounds from the HCP brain FC data when classifying gender from brain FC. (Zhang et al., 2018) showed in their study that regressing out the confounds, such as brain volume, in gender prediction with partial least squares regression did not affect the predictive performance much, and the important FC connections involved in the gender prediction and the brain volume prediction were generally different. (S. M. Smith et al., 2013) also discussed in their study that since males and females all have stronger and weaker connections in the important FC features involved in the gender prediction, which is also seen in our study, the gender classification is unlikely to be driven primarily by the gross group differences, such as the difference in brain volume. In our study each input connectivity matrix was also normalized to zero mean and unit variance with no gross asymmetry between the male and female groups.

## Conclusion

The present study successfully verified the feasibility of using DNN to predict gender from the resting state brain FC. The predictions under different scales of brain FC illustrate that the performance of DNN is highly related to the scale of brain FC, and the comparison between the nonlinear DNN model and the linear SVM model suggests that to achieve the best prediction result during model selection, the scale of the input brain FC should be taken into consideration. The proposed DNN feature ranking method provides a new understanding of the DNN model and manages to build the relationship between the input brain FC and the output gender. Bayesian DNN was also applied for the first time to a neuroscience classification problem, and showed how well the model uncertainty generated reacts to the uncertainty within the data. We believe that based on the success of the DNN and Bayesian DNN models in the FC gender prediction on the large scale HCP data, these models can be further applied in other fields of neuroscience.

## Acknowledgement

We gratefully acknowledge the Human Connectome Project for the data collection, data processing and data sharing. We also acknowledge the support of NVIDIA Corporation with the donation of the Titan Xp GPU used for this research. This work was supported by grants from the National Institutes of Health: U01-NS093650. The content is solely the responsibility of the authors and does not necessarily represent the official views of the National Institutes of Health.

# Supplementary Material

Table S1. The mean of the prediction accuracies across the 50 randomized cross validation permutations for each kind of predictive model and each kind of input. L – number of hidden layers in the DNN model; N – number of neurons in the first hidden layer of the DNN, the number of neurons in each of the rest hidden layers is half of this number.

|           | ICA25  | ICA50  | ICA100 | ICA200 | ICA300 | Template |
|-----------|--------|--------|--------|--------|--------|----------|
| L:1 N:20  | 0.8262 | 0.8790 | 0.9229 | 0.9373 | 0.9437 | 0.9146   |
| L:1 N:50  | 0.8260 | 0.8786 | 0.9235 | 0.9370 | 0.9434 | 0.9129   |
| L:1 N:100 | 0.8268 | 0.8783 | 0.9239 | 0.9369 | 0.9433 | 0.9130   |
| L:1 N:200 | 0.8281 | 0.8786 | 0.9237 | 0.9370 | 0.9427 | 0.9097   |
| L:2 N:20  | 0.8301 | 0.8767 | 0.9213 | 0.9371 | 0.9446 | 0.9144   |
| L:2 N:50  | 0.8318 | 0.8751 | 0.9205 | 0.9361 | 0.9432 | 0.9137   |
| L:2 N:100 | 0.8316 | 0.8764 | 0.9213 | 0.9350 | 0.9425 | 0.9134   |
| L:2 N:200 | 0.8314 | 0.8766 | 0.9215 | 0.9355 | 0.9418 | 0.9122   |
| L:3 N:20  | 0.8288 | 0.8761 | 0.9214 | 0.9368 | 0.9443 | 0.9142   |
| L:3 N:50  | 0.8300 | 0.8767 | 0.9198 | 0.9357 | 0.9434 | 0.9114   |
| L:3 N:100 | 0.8301 | 0.8770 | 0.9198 | 0.9350 | 0.9433 | 0.9126   |
| L:3 N:200 | 0.8300 | 0.8763 | 0.9202 | 0.9354 | 0.9414 | 0.9123   |
| SVM-linear| 0.8116 | 0.8717 | 0.9264 | 0.9411 | 0.9489 | 0.9214   |
| SVM-poly2 | 0.8144 | 0.8527 | 0.8956 | 0.9256 | 0.9311 | 0.8922   |

Table S2. The standard deviation of the prediction accuracies across the 50 randomized cross validation permutations for each kind of predictive model and each kind of input. L – number of hidden layers in the DNN model; N – number of neurons in the first hidden layer of the DNN, the number of neurons in each of the rest hidden layers is half of this number.

|           | ICA25  | ICA50  | ICA100 | ICA200 | ICA300 | Template |
|-----------|--------|--------|--------|--------|--------|----------|
| L:1 N:20  | 0.0094 | 0.0076 | 0.0068 | 0.0053 | 0.0061 | 0.0086   |
| L:1 N:50  | 0.0091 | 0.0077 | 0.0068 | 0.0060 | 0.0067 | 0.0079   |
| L:1 N:100 | 0.0091 | 0.0067 | 0.0062 | 0.0054 | 0.0061 | 0.0078   |
| L:1 N:200 | 0.0100 | 0.0076 | 0.0066 | 0.0060 | 0.0061 | 0.0084   |
| L:2 N:20  | 0.0101 | 0.0072 | 0.0075 | 0.0060 | 0.0070 | 0.0089   |
| L:2 N:50  | 0.0106 | 0.0078 | 0.0065 | 0.0062 | 0.0064 | 0.0078   |
| L:2 N:100 | 0.0106 | 0.0084 | 0.0067 | 0.0061 | 0.0063 | 0.0081   |
| L:2 N:200 | 0.0103 | 0.0074 | 0.0061 | 0.0056 | 0.0063 | 0.0080   |
| L:3 N:20  | 0.0094 | 0.0068 | 0.0067 | 0.0063 | 0.0069 | 0.0086   |
| L:3 N:50  | 0.0106 | 0.0069 | 0.0068 | 0.0050 | 0.0060 | 0.0080   |
| L:3 N:100 | 0.0100 | 0.0071 | 0.0071 | 0.0066 | 0.0061 | 0.0084   |
| L:3 N:200 | 0.0109 | 0.0070 | 0.0067 | 0.0053 | 0.0061 | 0.0081   |

| SVM-linear | 0.0102 | 0.0090 | 0.0074 | 0.0061 | 0.0069 | 0.0093 |
| SVM-poly2 | 0.0111 | 0.0109 | 0.0068 | 0.0057 | 0.0064 | 0.0087 |

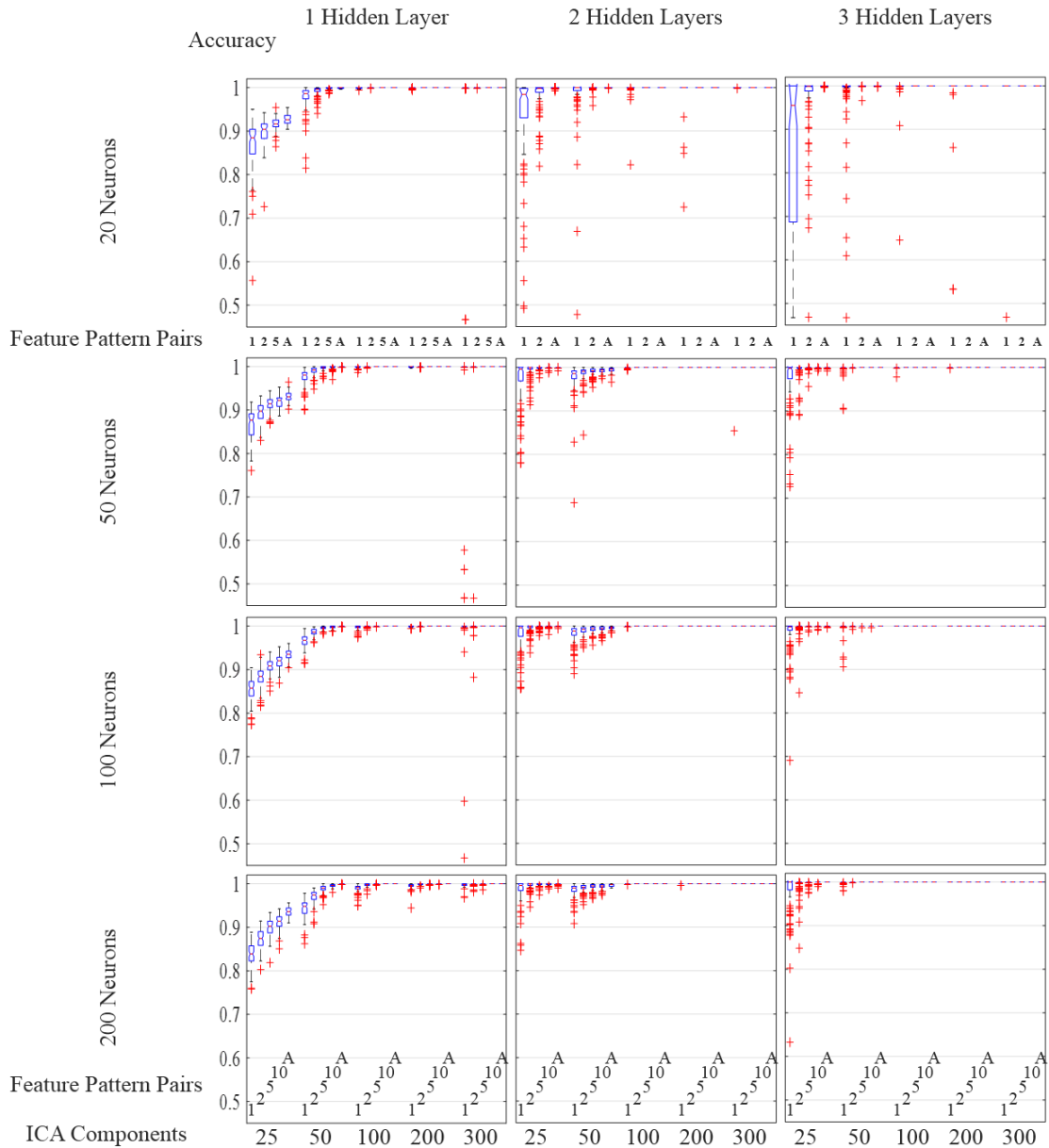

Fig. S1. Prediction accuracy recovered by the several most important high-level male and female feature pairs in the predictions on the training dataset. '1', '2', '5' and '10' mean that the predictions were made by the most important 1, 2, 5, 10 male and female feature pairs in the last hidden layer respectively. 'A' means the predictions were made by all the features..

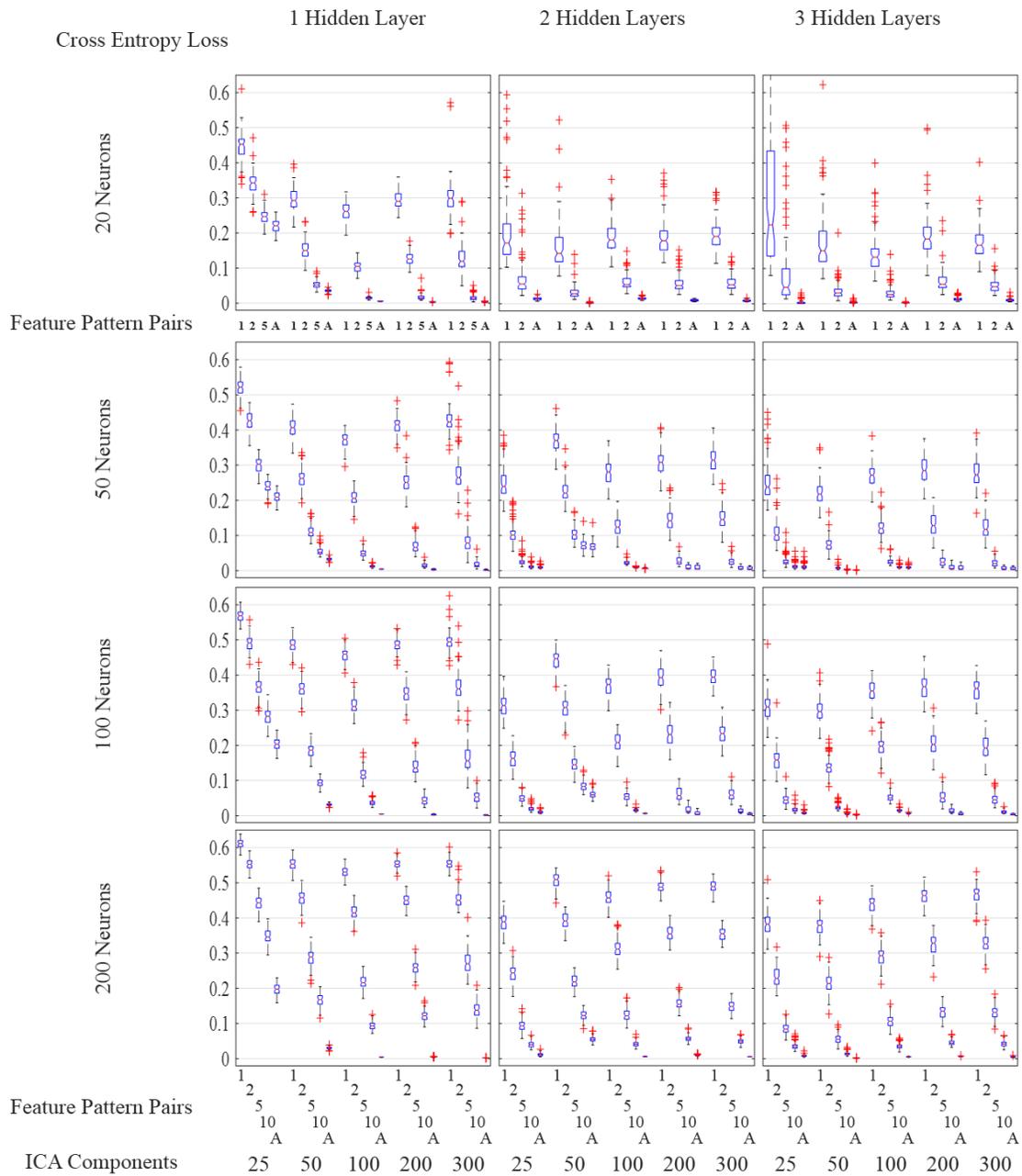

Fig. S2. The cross entropy loss achieved by the several most important high-level male and female feature pairs in the predictions on the training dataset. '1', '2', '5' and '10' mean that the predictions were made by the most important 1, 2, 5, 10 male and female feature pairs in the last hidden layer respectively. 'A' means the predictions were made by all the features.